\documentclass[aps,prb,twocolumn,showpacs,superscriptaddress,reprint,floatfix]{revtex4-1}

\usepackage{amsmath, amsfonts, amssymb}
\usepackage{enumerate}
\usepackage{graphicx}
\usepackage{color}
\usepackage[breaklinks]{hyperref}
\usepackage{paralist}
\usepackage[normalem]{ulem}

\newcommand{\eref}[1]{Eq.~(\ref{#1})}
\newcommand{\fref}[1]{Fig.~\ref{#1}}

\newcommand{\sref}[1]{Sec.~\ref{#1}}

\newcommand{\up}{\uparrow}
\newcommand{\dw}{\downarrow}
\newcommand{\e}{\mathrm{e}}

\newcommand{\si}{\hat{\sigma}_{0}}
\newcommand{\sx}{\hat{\sigma}_{1}}
\newcommand{\sy}{\hat{\sigma}_{2}}
\newcommand{\sz}{\hat{\sigma}_{3}}

\newcommand{\tx}{\hat{\tau}_{1}}
\newcommand{\ty}{\hat{\tau}_{2}}
\newcommand{\tz}{\hat{\tau}_{3}}

\newcommand{\mean}[1]{\langle #1 \rangle}

\newcommand{\mbf}[1]{\mathbf{ #1 }}

\DeclareMathOperator{\imag}{Im}

\DeclareMathOperator{\tr}{Tr}

\begin{document}


\title{Tunable hybridization of Majorana bound states at the quantum spin Hall edge} 

\newcommand{\wuerzburg}{Institute for Theoretical Physics and Astrophysics,
	University of W\"{u}rzburg, D-97074 W\"{u}rzburg, Germany}
\newcommand{\aalto}{Department of Applied Physics,
	Aalto University, FIN-00076 Aalto, Finland}

\author{Felix Keidel}
\affiliation{\wuerzburg}

\author{Pablo Burset}
\affiliation{\aalto}

\author{Bj\"orn Trauzettel}
\affiliation{\wuerzburg}
 
\date{\today}



\begin{abstract}
Confinement at the helical edge of a topological insulator is possible in the presence of proximity-induced magnetic (F) or superconducting (S) order. The interplay of both phenomena leads to the formation of localized Majorana bound states (MBS) or likewise (under certain resonance conditions) the formation of ordinary Andreev bound states (ABS).
We investigate the properties of bound states in junctions composed of alternating regions of F or S barriers. Interestingly, the direction of magnetization in F regions and the relative superconducting phase between S regions can be exploited to hybridize MBS or ABS at will. We show that the local properties of MBS translate into a particular nonlocal superconducting pairing amplitude. Remarkably, the symmetry of the pairing amplitude contains information about the nature of the bound state that it stems from. Hence, this symmetry can in principle be used to distinguish MBS from ABS, owing to the strong connection between local density of states and nonlocal pairing in our setup. 
\end{abstract}

\maketitle

\section{Introduction \label{sec:intro}} 
Majorana bound states (MBS) are localized zero energy excitations at the edges of topological superconductors that are in the focus of recent condensed matter research\cite{Alicea_Majorana,Beenakker_Majorana}. This strong interest is mainly due to their non-Abelian statistics, which makes them ideal building blocks for topological quantum computers\cite{Kitaev-anyons,Nayak_RMP}.
Notably, MBS can be engineered by combining conventional $s$-wave superconductors with materials featuring strong spin-orbit coupling. The underlying idea is quite similar for most proposals: $s$-wave superconducting pairing plus spin-orbit coupling leads to an effective $p$-wave pairing that is usually more pronounced if some kind of magnetic factor (e.g., a magnetic field or magnetic adatoms) is added to the system.
The most prominent example thereof is based on semiconducting nanowires\cite{Oreg_2010,Sau_2010,Lutchyn_2010,Alicea_2010} in the presence of an external magnetic field and proximity-induced superconductivity. This concrete proposal has been tested in several ground-breaking experiments, which provided evidence for the emergence of MBS in the form of zero bias conductance peaks\cite{Mourik_2012,Rokhinson_2012,Das_2012}.

A somewhat more robust platform to find MBS has been predicted on the basis of the helical edge states of quantum spin Hall insulators (QSHI) \cite{QSHI_2007,Brune_2010,Sullivan_2011,Brune_2012}. Since these helical edge states form Dirac fermions in one spatial dimension (1D), they can be viewed as an electronic system mimicking infinitely strong spin-orbit coupling. In the presence of barriers, based on proximity-induced $s$-wave superconductivity and ferromagnetism, MBS emerge at the helical edge \cite{FuKane_2009,Crepin_2014} which are stable against the influence of Coulomb interaction \cite{Crepin_PRL2014}. From the experimental side, superconductivity has been proximity-induced into QSHI \cite{Sullivan_2012,Hart_2014}. The evidence for MBS in this realization of a topological superconductor has culminated, so far, in experiments on (missing) Shapiro steps in Josephson junctions based on helical edge states \cite{Bocquillon_2017}. Importantly for the physics discussed in the following, the emergence of MBS at the helical edge can be connected to particular transport signatures, like crossed Andreev reflections\cite{Nilsson_2008} that are intimately connected to the presence of odd-frequency superconductivity \cite{Crepin_2015}.

In this article, we study the physical properties of alternating F-S and/or S-S cavities that introduce confinement into helical liquids. The main objectives are: (i) a better understanding of the tunability of hybridization of MBS in neighbouring cavities and (ii) a classification of the nonlocal pairing amplitudes which are, surprisingly, related to the local density of states in these hybrid structures. We show below that the two objectives are strongly connected to each other. Therefore, they help us to develop a deep understanding of the underlying physics.
Due to the topological origin of the helical edge state, the number of bound states formed within the F-S cavity is always odd. Independently of the cavity's dimensions, at least one MBS is naturally emerging.
Additional (trivial) Andreev bound states (ABS) may appear but they come in pairs at finite energies. While the MBS is robust (i.e., pinned to zero excitation energy) against variations of system parameters, the ABS are not.
Two consecutive cavities in a F-S-F setup feature a rich hybridization between the bound states of each individual cavity (see \fref{fig:sketch} below). We find, for instance, a condition for destructive interference of excitations forming the bound states that totally removes the hybridization of cavity states. With application potential in topological quantum computing, this understanding allows to (de)couple neighboring MBS at will. In this F-S-F setup, the hybridization can, for instance, be tuned by the relative angle of magnetization of the two outer F barriers. Likewise, an S-F-S setup can be used for the same purpose where now the hybridization can be tuned by the relative superconducting phase of the two outer S barriers. The story gets a twist in all-S setups (the S-S-S combination) where helical MBS can be formed under certain combinations of relative superconducting phases of the three S barriers. In that latter scenario, the tunability of the hybridization of helical MBS formed in neighboring S-S cavities is unfortunately less pronounced than in the cases where F barriers are included. Nevertheless, this analysis is important to identify the ideal Majorana platform in which MBS can be (de)coupled on demand.


\begin{figure}
	\includegraphics[width=1.0\columnwidth]{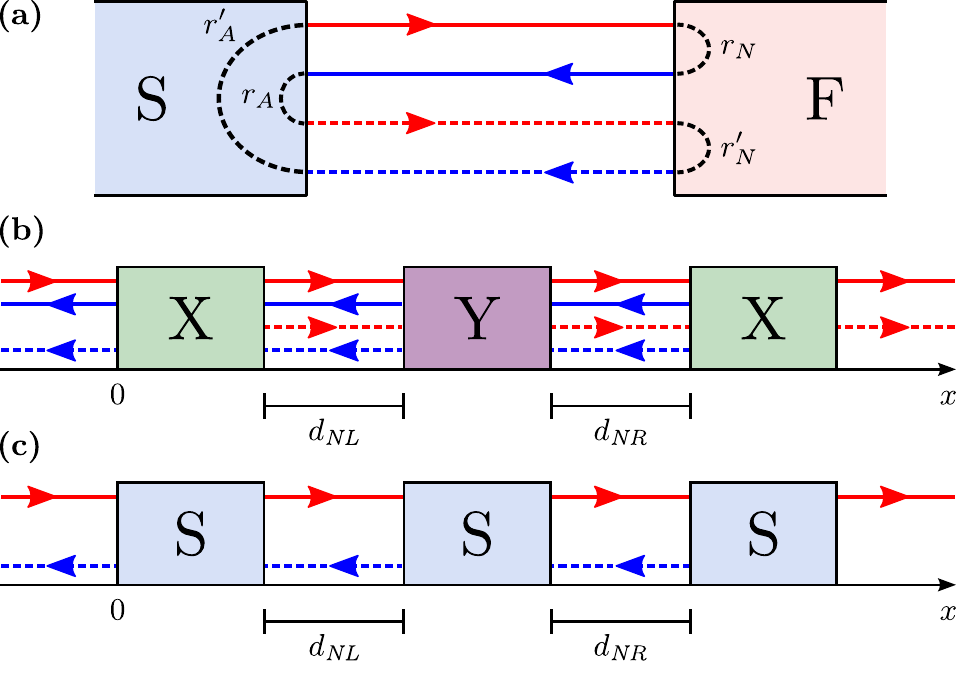}
	\caption{\label{fig:sketch} 
		Confinement at the helical edge of a topological insulator. 
		(a) Basic scattering processes and formation of bound states at the S-F cavity. (b) General sketch of a double cavity X-Y-X, where X and Y can be either S or F regions. Right moving spin-up electrons (holes) are indicated by solid (dashed) red arrows. Blue arrows indicate equivalent left-moving spin-down particles. In the text, we label outer normal regions (leads) L and R and the intermediate ones NL and NR. The barriers are named XL, Y, and XR. To avoid ambiguity, physical parameters of different regions are labeled accordingly. (c) Setup with three superconductors. Here, helicity forbids transmission of holes through the junction. Normal regions are labeled as in (b), whereas the S barriers are referred to as SL, SM, and SR.}
\end{figure}


Interestingly, the emergence of MBS and ABS in these hybrid structures is strongly connected to a particular nonlocal superconducting pairing. The physics reason behind is related to the helicity of the edge states. Helicity is highly restrictive for the allowed superconducting pairing channels because it implies a broken spin rotation invariance.  In the presence of such hybridized MBS, e.g., in the S-F-S or the F-S-F setup mentioned above, the induced pairing from the superconductor can be nonlocal, meaning that Cooper pairs are formed from electronic states belonging to different cavities and or even different leads. Owing to the perfect spin-momentum locking of the helical edge, such nonlocal Cooper pairs can only be in a spin-polarized triplet state, with total suppression of the nonlocal spin singlet state.
We demonstrate below that odd-frequency equal-spin triplet pairing emerges in a particular way precisely at the energies where bound states (MBS or ABS) are formed. Notably, it is dominant for the zero-energy MBS which indicates a strong connection between odd-frequency pairing and the emergence of MBS in these nanostructures\cite{Asano_2013}. 
Likewise in the S-S-S setup, we identify the appearance of unconventional pairing. In that case, it is of the odd-frequency singlet and odd-frequency triplet (with zero $z$-component) type. 
Odd-frequency pairings can be dominant on resonance, but their magnitude is usually comparable to the even-frequency ones.

The article is organized as follows. We introduce the formalism and describe the formation of bound states and their hybridization in \sref{sec:model}. We discuss the tunable hybridization in \sref{sec:hyb} and the unconventional superconducting pairing in \sref{sec:pair}. We conclude in \sref{sec:conc}.


\section{Model  \label{sec:model} }
\subsection{Formation of bound states at the helical edge}
A helical liquid is formed by a pair of edge states of a QSHI with locked group velocity and spin orientation. We consider the spin quantization axis to be along the $z$ direction and describe our system using the Nambu spinor $\Psi\!=\!(\psi_{R\uparrow}, \psi_{L\downarrow}, \psi^\dagger_{L\downarrow}, - \psi^\dagger_{R\uparrow})^T$, where the electron field operators $\psi^\dagger_{R\uparrow}$ and $\psi^\dagger_{L\downarrow}$ ($\psi_{R\uparrow}$ and $\psi_{L\downarrow}$) create (annihilate) right movers with spin $\up$ and left movers with opposite spin, respectively. 
In the following, the helical liquid is assumed to be in proximity to several superconducting and ferromagnetic electrodes, inducing either singlet $s$-wave superconductivity or Zeeman coupling in the edge states. 
The Bogoliubov-de Gennes Hamiltonian describing the system takes the form (we set $\hbar\!=\!1$ and the Fermi velocity $v_F\!=\!1$)
\begin{equation} \label{eq:hamil}
\mathcal{H}_\text{BdG}(x)=  p_x \sz \tz + \mathbf{m}(x) \cdot \boldsymbol{\sigma} + \boldsymbol{\Delta}(x) \cdot \boldsymbol{\tau} ,
\end{equation}
where $p_x\!\!=\!\!-i\partial_x$, and the Pauli matrices $\boldsymbol{\sigma}\!=\!(\sx,\sy,\sz)$ [$\boldsymbol{\tau}\!=\!(\tx,\ty,\tz)$] act in spin (particle-hole) space. 
To illustrate the general effect that the superconducting and Zeeman couplings have on the helical states, we define $\boldsymbol{\Delta}(x) \!=\! [ \Delta_0\cos \chi,\Delta_0\sin \chi , -\mu]^T$ and $\boldsymbol{m}(x) \!=\! [ m_{\parallel} \cos\phi,m_{\parallel}\sin\phi , m_z]^T$. 
The notation in \eref{eq:hamil} allows us to identify the effect that the components of $\boldsymbol{\Delta}$ and $\boldsymbol{m}$ have on the system. 
For example, the chemical potential $\mu$ and the perpendicular component of the magnetization $m_z$ only induce a renormalization of the momentum. In this article, we distinguish among normal state regions (N) with $\Delta_0 \! = \! 0$ and $m_\parallel \! = \! m_z \! = \! 0$, S regions with $\Delta_0 \! \neq \! 0$ and $m_\parallel \! = \! m_z \! = \! 0$, and F domains with $\Delta_0 \!=\!0$ and $m_\parallel, m_z \neq 0$. We impose rigid boundary conditions such that $\boldsymbol{\Delta}(x)$ and $\boldsymbol{m}(x)$ remain constant within each region.

Interestingly, the pairing potential $\Delta_0 e^{i\chi}$, with $\chi$ the superconducting phase, and the in-plane magnetization $m_{\parallel}\e^{i\phi}$, with magnitude $m_{\parallel}$ and orientation angle $\phi$, open a gap in the spectrum of the helical liquid. 
The former couples electrons and holes in S regions, while the latter connects particles with opposite spins in F regions. 
The opening of a gap results in the otherwise forbidden back-scattering at the interface with an N domain, i.e. Andreev reflections take place at NS interfaces and spin-flip normal reflections occur at NF junctions (see \fref{fig:sketch}(a)). 
A resonant state can thus be formed between two regions of the same type \cite{Timm_2012,Sassetti_2013,Traverso_2013,Crepin_2014}. For an F-F cavity, where we omit the intermediate N region for brevity of notation, the resonance condition for bound state energies $\omega$ reads as
\begin{equation} \label{eq:ff-cav}
2\pi n = 2\omega d_N \pm 2\mu_N d_N \pm\Delta\phi - 2\arccos\left(\frac{\omega \pm \mu_F}{m_{\parallel}}\right),
\end{equation}
with $n$ an integer, $d_N$ the cavity's width, $\mu_N$ the chemical potential in the intermediate normal region and $\Delta\phi$ the relative angle of the in-plane magnetizations in each region (with $\pm$ for electrons and holes, respectively). For simplicity we have assumed that the magnitude of the in-plane magnetization, $m_{\parallel}$, and the chemical potential, $\mu_F$, are the same for both ferromagnetic regions. 

Similarly, the S-S cavity, or Josephson junction if the superconductors are semi-infinite, features bound states at the energies given by 
\begin{equation}\label{eq:ss-cav}
2 \pi n = 2 \omega d_N- 2\arccos(\omega/\Delta_0)\pm\Delta\chi ,
\end{equation}
where we have assumed that the superconducting gap is the same for both S regions and $\Delta\chi$ is the phase difference. Note that since the bound states are composed out of electrons \textit{and} holes, this expression is now independent of $\mu_N$. Furthermore, the $z$-component of the magnetic field and the chemical potential in the superconductors do not appear in \eref{eq:ff-cav} and \eref{eq:ss-cav}, respectively, since the particles do not enter the semi-infinite barriers.

The bound states\footnotemark[1] formed in hybrid S-F and F-S junctions differ from the previous resonances in two important aspects\cite{Crepin_2014}. They involve all four types of particles in our system, namely, right-moving spin-up and left-moving spin-down electrons and holes. Moreover, since they require two Andreev and two normal reflections, \textit{cf.} \fref{fig:sketch}(a), the distance traveled by the standing wave is four times the width of the intermediate normal region, twice the usual distance. Assuming semi-infinite S and F regions, we find the bound state condition
\begin{gather}
2\pi n= 4\omega d_N - 2\arccos\left(\frac{\omega}{\Delta_0}\right) \nonumber \\ -\arccos\left(\frac{\omega-\mu_F}{m_{\parallel}}\right) - \arccos\left(\frac{\omega+\mu_F}{m_{\parallel}}\right)  , 
\label{eq:sf-cav}
\end{gather}
with $\mu_F$ the chemical potential in the F region. 
Contrary to the usual Fabry-P\'erot cavity, \eref{eq:sf-cav} always has a solution for $\omega\!=\!0$ in addition to potential pairs of states at finite energies. Therefore, the total number of states is always odd. The extra zero-energy state is a topologically protected Majorana bound state (MBS) \cite{Crepin_2014}. We refer to the Fabry-P\'erot bound states at non-zero energies as Andreev bound states (ABS). 

\footnotetext[1]{Only in the particular case where the F or S regions extend to infinity one can speak of a bound state. In this work we consider finite S and F regions, but we still refer to the resonant states in the F-S cavities as bound states since they are mostly confined and appear for well-defined energies with a controllable width. }

\subsection{Correlation functions}
To analyze the induced pairing and bound states in our system, we make use of the retarded Green function defined as
\begin{equation}
G^R(x,t;x',t') = -i \theta(t-t') \, \langle \lbrace \Psi(x,t), \Psi^\dagger(x',t')  \rbrace \rangle ,
\end{equation}
where $x,x'$ and $t,t'$ denote position and time coordinates, respectively. In the following, we work in the frequency representation by using the Fourier transform $G^R(x,x',\omega) \!=\! \int \! \mathrm{d}t \, G^R(x,t;x',t') \e^{i (\omega + i 0^+ ) (t-t')}$. We construct the Green function from the asymptotic form of the scattering states of the system, following the method described in Refs.~\onlinecite{Kashiwaya_2000,Herrera_2010,Crepin_2015,Burset_2015}. 
Particle-hole symmetry of the Hamiltonian in \eref{eq:hamil}, i.e., $U_\text{C}\mathcal{H}^*_\text{BdG}U_\text{C}^{\dagger}\!=\!-\mathcal{H}_\text{BdG}$, with $U_\text{C}\!=\!\sy\ty$, yields a connection between different Nambu (particle-hole) sectors of the Green's function. Indeed, we find that 
\begin{align*}
G_{ee}^R(x,x',-\omega)={}& -\sy G_{hh}^R(x,x',\omega)^*\sy , \\ 
G_{eh}^R(x,x',-\omega)={}& \sy G_{he}^R(x,x',\omega)^*\sy . 
\end{align*}
Therefore, to characterize our system we only need to specify the normal Green function, $G_{ee}^R(x,x',\omega)$, and the anomalous function $G_{eh}^R(x,x',\omega)$. 
From the normal Green function we obtain the local density of states (LDOS) as
\begin{equation}\label{eq:ldos}
\rho(x, \omega) = -\frac{1}{\pi} \imag \left[ \tr  G^R_{ee}(x,x,\omega)  \right].
\end{equation}
In the basis of Nambu spinors that we are using, $\Psi\!=\!(\psi_{R\uparrow}, \psi_{L\downarrow}, \psi^\dagger_{L\downarrow}, - \psi^\dagger_{R\uparrow})^T$, the anomalous Green function can be decomposed as follows
\begin{equation}\label{eq:anom}
G_{eh}^R(x,x',\omega)= f_0^R(x,x',\omega)\si + \sum\limits_{j=1}^{3} f_j^R(x,x',\omega) \hat{\sigma}_j ,
\end{equation}
with $f_0^R(x,x',\omega)$ the singlet (S) and $f_j^R(x,x',\omega)$ the triplet (T) components of the pairing amplitude.

We can now characterize the pairing amplitude's symmetry with respect to position and spin. 
First, we notice that at the helical edge of a QSHI, the strong spin-orbit locking imposes some restrictions on the anomalous correlators, i.e., on the formation of some type of Cooper pairs. Indeed, $\mean{\psi_{L\dw}(t)\psi_{R\dw}(0)}$ and $\mean{\psi_{R\up}(0)\psi_{L\up}(t)}$, with $t\!>\!0$, must vanish since they describe Cooper pairs formed by spin-down electrons moving rightwards and spin-up electrons moving leftwards; both forbidden by helicity. 
In Appendix \ref{app:sym-f}, we demonstrate how these restrictions affect the general symmetries of the anomalous Green functions. 
To define an anomalous Green function that fulfills Fermi-Dirac statistics, i.e., that is odd under the exchange of the two fermion operators, we need a combination of retarded and advanced Green functions as $f^R+f^A$, with $f^A$ the advanced anomalous Green function. From the general property $G^R(x,x',\omega)\!=\!G^A(x',x,\omega)^{\dagger}$, and using particle-hole symmetry, we find that the advanced anomalous Green function is written in terms of the retarded one as $[G_{eh}^A]_{\sigma\sigma'}(x,x',\omega)\!=\!-[G_{eh}^R]_{\sigma'\sigma}(x',x,-\omega)$, with $\sigma,\sigma'\!=\up,\dw$ the spin indices. 
Notice that the change in spin indices will only affect the singlet state, which is odd under that exchange. 

To fully identify the symmetry of the anomalous Green functions of \eref{eq:anom}, we consider symmetric and anti-symmetric superpositions of spatial coordinates as $f_{\mu,\pm}^R(\omega)\!=\![f_{\mu}^R(x,x',\omega)\!\pm\!f_{\mu}^R(x',x,\omega)]/2$, where $\mu\!=\!0$ for the singlet and $\mu\!=\!1,2,3$ for the triplet components. We can thus define the fully symmetrized anomalous functions
\begin{equation}\label{eq:anom-sym}
f^R_{\mu,\pm}(\omega) \!+\! f^A_{\mu,\pm}(\omega)=\!\! \left\{\! \begin{array}{cl}
f^R_{0,\pm}(\omega) \pm f^R_{0,\pm}(-\omega) & \text{(singlet)} \\ 
f^R_{j,\pm}(\omega) \mp f^R_{j,\pm}(-\omega) & \text{(triplet)} 
\end{array}\!\right. 
\end{equation}
As a result, the singlet components must be even (E) or odd (O) in both frequency and spatial coordinates at the same time. We label these pairing amplitudes ESE (even-frequency, singlet, even-parity) and OSO (odd-frequency, singlet, odd-parity). 
For the triplet states, the pairing amplitudes must have different symmetry with respect to energy and position. We thus have ETO (even-frequency, triplet, odd-parity) and OTE (odd-frequency, triplet, even-parity) pairings \cite{Linder_review,[{For more information about the classification of superconducting pairing amplitudes see }] Tanaka_JPSJ,*Eschrig_2010,*Eschrig_Review}. 

In what follows, we consider scattering at junctions where the leftmost and rightmost regions are in the normal state, \textit{cf}. \fref{fig:sketch}(b,c). Between those regions, we analyze a double F-S cavity in \sref{sec:fs} (F-S-F and S-F-S) or the all-electrical double S-S cavity in \sref{sec:sss}. For all cases, we can provide analytical expressions for the pairing amplitudes evaluated at the outermost normal regions as a function of the scattering amplitudes. Notice that the scattering amplitudes change for each setup, but the structure of the pairing amplitude is the same. 
The only non-zero pairing amplitudes with spatial coordinates $x_{L}$ in the leftmost region and $x_{R}\!>\!x_{L}$ in the rightmost are given by the equal-spin components
\begin{widetext}
	\begin{subequations}\label{eq:FLR}
\begin{align}
		f^{+}_{\text{ETO},\text{OTE}}(\omega,x_{R},x_{L})={}&\frac{-i}{2v_F}\e^{i\mu_{L} x_{L}+i \mu_{R} x_{R}} \left[ \e^{-i\omega(x_{L}-x_{R})} t_\mathit{hL,eR}(\omega) \pm \e^{i\omega(x_{L}-x_{R})} t_\mathit{hL,eR}(-\omega) \right], \\
		f^{-}_{\text{ETO},\text{OTE}}(\omega,x_{L},x_{R})={}&\frac{\pm i}{2v_F} \e^{-i\mu_{L} x_{L}-i \mu_{R} x_{R}} \left[ \e^{i\omega(x_{L}-x_{R})} t_\mathit{hR,eL}(\omega) \pm \e^{-i\omega(x_{L}-x_{R})} t_\mathit{hR,eL}(-\omega) \right], 
\end{align}
	\end{subequations}
\end{widetext}
where the indices $\pm$ denote the spin-polarized triplet states $\up\up$ and $\dw\dw$, respectively.

\begin{figure}
	\includegraphics[width=\columnwidth]{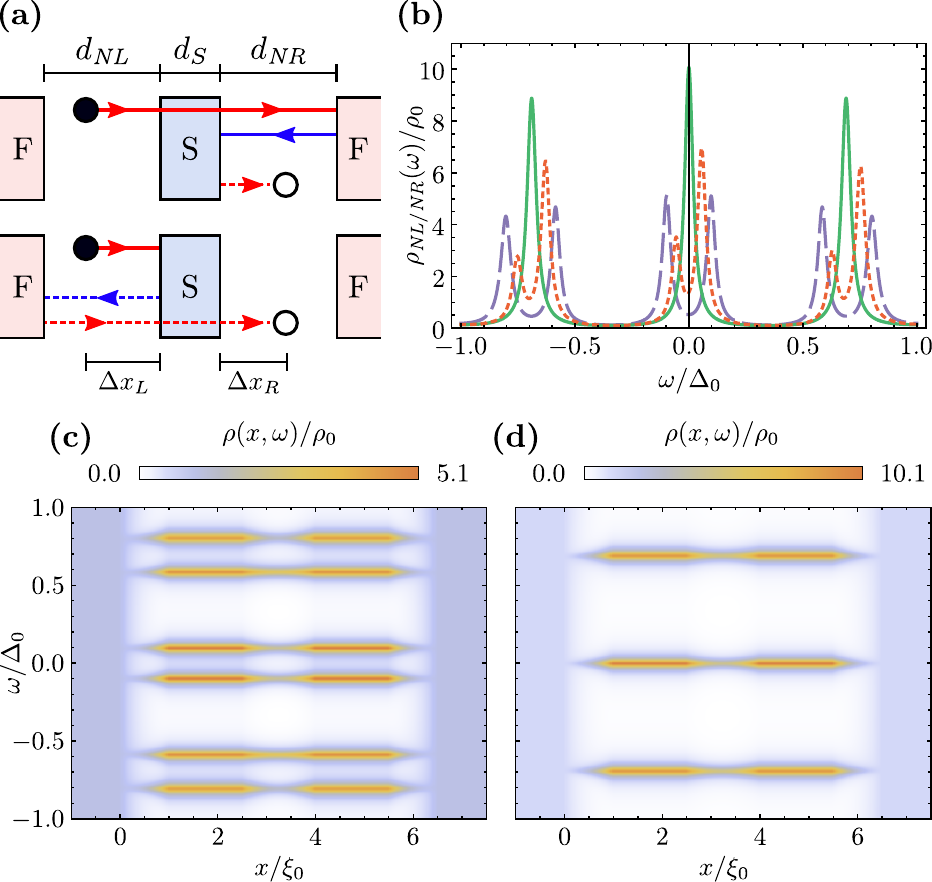}
	\caption{\label{fig:nonlocal} Local and nonlocal bound states. 
		(a) Schematic representation of two possible paths for an electron to scatter across the S region and end up as a hole. Both processes only differ by a phase (due to equal number of tunnelings through barriers), and can destructively interfere if the resonance condition is met. Plot of the LDOS in dependence of energy (b) for $\phi_{R}=0$ (dashed line), $0.6\pi$ (dotted line) and $\pi$ (solid line). (c,d) Density plot of the LDOS as a function of the energy for full (c)/without (d) hybridization. We take $m_{\parallel}\!=\!1.5\Delta_0$ as field strength and widths $d_{\mathit{NL}}\!=\!d_{\mathit{NR}}\!=\!d_\mathit{S}=1.5\xi_0$, $d_{\mathit{FL}}\!=\!d_{\mathit{FR}}\!=\!1\xi_0$ for all plots. 
        }
\end{figure}

The amplitude $t_\mathit{hL,eR}$ ($t_\mathit{hR,eL}$) corresponds to the scattering process in which a hole impinging from the leftmost (rightmost) normal region is transmitted as an electron to the rightmost (leftmost) one -- a crossed Andreev reflection (CAR). 
As a result, nonlocal pairing is directly connected to crossed Andreev processes and will vanish if this transport channel is closed. 
Helicity of the edge state of the QSHI yields that crossed Andreev processes involve particles with the same spin \cite{Law_2014,Crepin_2015}. 
From \eref{eq:FLR}, we immediately see that nonlocal Cooper pairs with electrons going to different edges of the junction can only be in a spin-polarized triplet state.


\section{Tunable hybridization of bound states between cavities  \label{sec:hyb}}
In all the double cavities considered here, there is a possibility for the bound states from different cavities to hybridize due to the finite width of the central barrier, be it S or F. 
The hybridization of the quasi-bound states originates from possible scattering events of quasiparticle excitations between the intermediate normal domains, \textit{cf}. \fref{fig:nonlocal}(a). 
By tuning several system parameters, one can effectively achieve destructive interference and prohibit quasiparticle transport between those N regions. 
For example, in a F-S-F junction with semi-infinite F regions, without loss of generality, we can set the chemical potential in the ferromagnets to zero ($\mu_{\mathit{FL}}\!=\!\mu_\mathit{FR}\!=\!0$) and assume a symmetric setup with $m_{\parallel,L}\!=\!m_{\parallel,R}$ and $d_\mathit{NL}\!=\!d_\mathit{NR}$, with $d_\mathit{NL,NR}$ the width of each cavity. 
The condition for destructive interference is then given by
\begin{equation} \label{eq:int-fsf}
\Delta\phi/2 + \mu_\mathit{NL} d_\mathit{NL} + \mu_{S} d_{S} + \mu_\mathit{NR} d_\mathit{NR} = (n+1/2)\pi ,
\end{equation}
with $\Delta\phi\!=\!\phi_R\!-\!\phi_L$ the relative orientation of the magnetizations in each F region, and $d_\mathit{NL,S,NR}$ ($\mu_\mathit{NL,S,NR}$) the width (chemical potential) of the intermediate N and S regions. 
Analogously, destructive interference occurs at the S-F-S double cavity if
\begin{equation} \label{eq:int-sfs}
\Delta\chi - m_{z} d_{F} = (n+1/2)\pi ,
\end{equation}
with $d_F$ the width of the central F region, $\Delta\chi\!=\!\chi_R\!-\!\chi_L$ the superconducting phase difference and where we have set $\Delta_{L}\!=\!\Delta_{R}\!\equiv\!\Delta_0$. 
For cavities with different widths, Eqs. (\ref{eq:int-fsf}) and (\ref{eq:int-sfs}) acquire an extra phase $(d_\mathit{NL}-d_\mathit{NR})\omega$. However, the MBS with $\omega\!=\!0$ still features the same destructive interference. 
This is an important distinction between MBS and finite energy ABS with consequences further explored below. 

In both cases, hybridization between each cavity's levels can be tuned by adjusting the chemical potentials in the intermediate regions, which can be done by local gates. Additionally, for the F-S-F setup, hybridization is controlled by the relative orientation of the in-plane magnetization at each F region. The phase difference between superconductors plays the same role for the S-F-S junction. 
The splitting of cavity levels by hybridization is always the same, independently of which parameters are used. One can thus fix all parameters (local chemical potentials and magnetic orientations or superconducting phases) and tune the hybridization at will using only one parameter, which we call \textit{hybridization phase} from now on. 

The LDOS at the intermediate normal regions, as defined in \eref{eq:ldos}, directly explores the hybridization of the bound states. 
In the absence of hybridization, bound states form at each intermediate region. As a result, the LDOS always features a peak at zero energy (MBS) and, if the conditions given by Eqs. (\ref{eq:ff-cav},\ref{eq:ss-cav},\ref{eq:sf-cav}) are fulfilled, at finite energies (ABS). The solid line of \fref{fig:nonlocal}(b) shows these non-hybridized states. 
When the hybridization phase is such that destructive interference occurs between the intermediate left and right normal regions, the bound states split. This case is illustrated by the dotted and dashed lines of \fref{fig:nonlocal}(b). 
The spatially resolved LDOS across the junction, \fref{fig:nonlocal}(c,d), shows how the states are mainly localized in the cavities and connected through the finite central region. Note that throughout this article we normalize the LDOS to the constant LDOS within the leads, $\rho_0$. 

Hybridization happens to all states, including the MBS, and has an important consequence on the transport properties of the junction. 
For a symmetric double cavity ($d_\mathit{NL}\!=\!d_\mathit{NR}$), when we consider injection of electrons from the leftmost normal region, in the absence of hybridization we only find transmitted electrons on the rightmost region (electron co-tunneling processes). 
Notice that this case is still different from a helical NSN junction\cite{Adroguer_2010,Reinthaler_2013}, since transmission is only possible for the energies of the bound states (on resonance). 
However, when the bound states at each cavity hybridize, nonlocal (crossed) Andreev reflections are possible. The effect of hybridization thus strongly affects the nonlocal conductance of the junction. 
For the asymmetric double cavity ($d_\mathit{NL}\!\neq\!d_\mathit{NR}$), the number of ABSs on each cavity can be different. In that case, the levels do not hybridize and crossed Andreev reflection takes place on resonance \cite{Crepin_2015}. 

\begin{figure*}
\includegraphics[width=\textwidth]{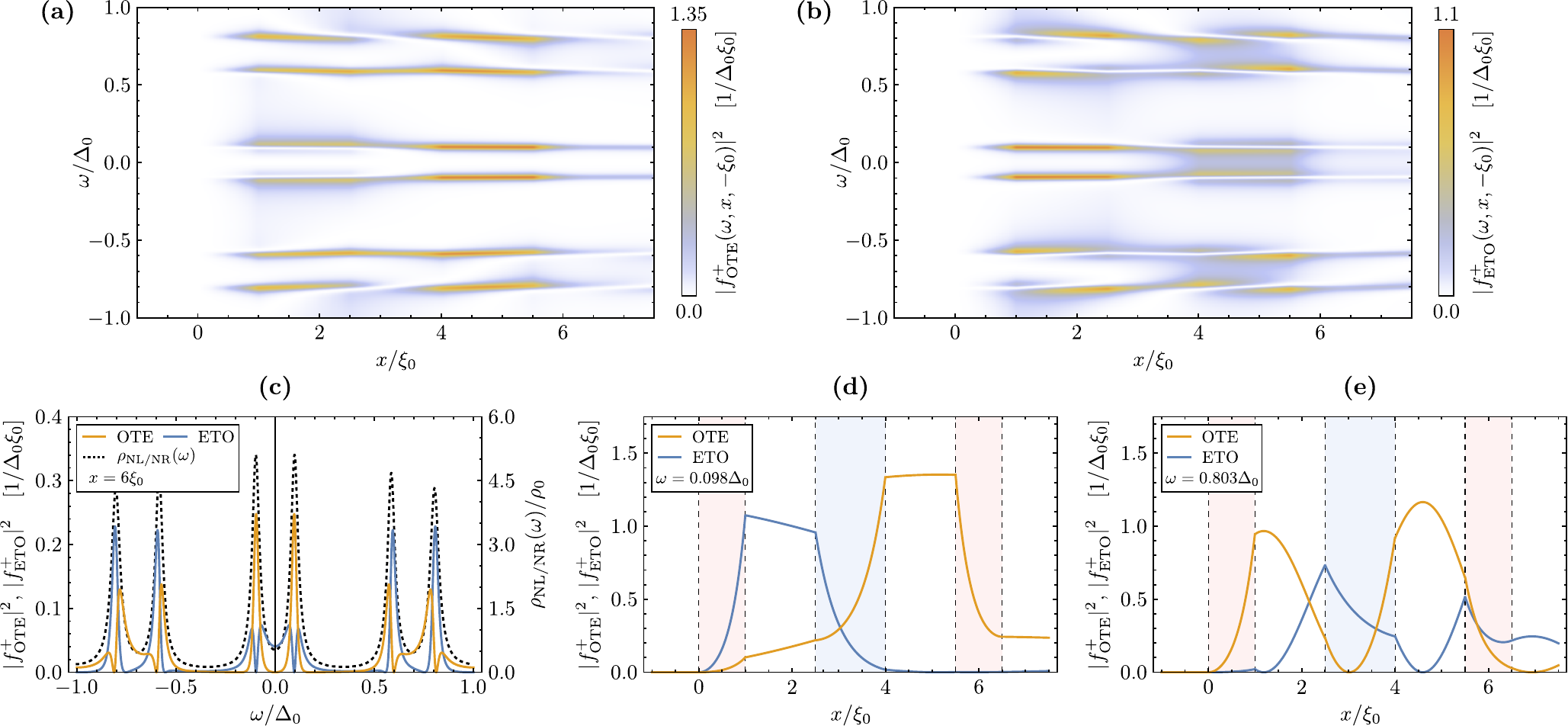}
	\caption{\label{fig:results1} 
		Local density of states and pairing amplitude for the F-S-F setup. 
		(a,b) Maps of $\mathrm{OTE}_+$ (a) and $\mathrm{ETO}_+$ (b) pairing as a function of the energy and position $x$, with fixed $x'\!=\!-1.5\xi_0$ in the left lead. 
		(c) For fixed position $x\!=\!6\xi_0$, plot of the pairings and LDOS as a function of the energy. 
		(d,e) Plot of the pairings as a function of the position for fixed energies $\omega\!=\!0.098 \Delta_0$ (d) and $\omega\!=\!0.803\Delta_0$ (e). }
\end{figure*}

\begin{figure}
\includegraphics[width=\columnwidth]{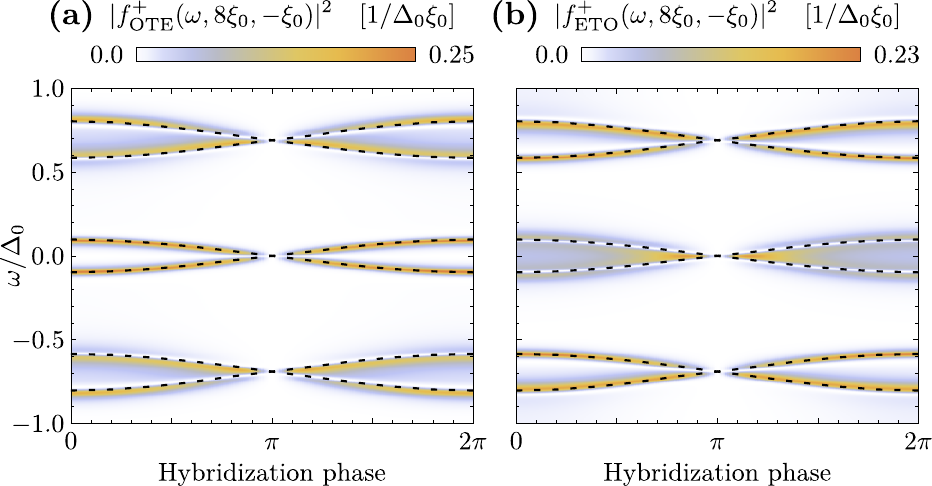}
\caption{\label{fig:hyb} Connection between induced nonlocal pairing and local density of states.  $\mathrm{OTE}_+$ (a) and $\mathrm{ETO}_+$ (b) pairing amplitudes for $x' \! = \! -\xi_0$ and $x \! = \! 8\xi_0$ as a function of the hybridization phase and energy. The dashed, black lines represent the maxima of the LDOS within the cavities, $\rho_\mathit{NL/NR}(\omega)$. }
\end{figure}


\section{Proximity-induced pairing in a double cavity \label{sec:pair}}
\subsection{SFS and FSF setup \label{sec:fs}}
Interestingly, the properties of the localized quasi-bound states described above are also reflected in the induced anomalous nonlocal correlations in the system. 
To illustrate this claim, we use \eref{eq:FLR} to study the even (ETO) and odd-frequency (OTE) parts of the equal-spin triplet pairing. 
The reason why equal-spin pairing is a natural way to capture the effect of bound states on nonlocal anomalous correlations is twofold: 
\begin{inparaenum}[(i)]
\item due to helicity restrictions, nonlocal, anomalous correlations across the whole scattering region can only involve equal spin particles and holes and
\item equal-spin pairing requires the mixing of electrons and holes of both spin channels resulting from a combination of S {\it and} F regions. 
\end{inparaenum}

We start with the F-S-F structure introduced in \sref{sec:hyb} and consider fully hybridized states at first. To study the induced nonlocal pairing, we fix one spatial argument of the correlation function in the leftmost lead ($x' \! = \! -\xi_0$, with $\xi_0\!=\! v_F/\Delta_0$ the superconducting coherence length) and sweep the other across the junction. This represents processes that take spin up particles or holes from the left lead to the right lead; we thus focus on the $\up\up$-component of the pairing. 
In the upper panels of \fref{fig:results1}, we plot the position $x$ and energy $\omega$ dependence of $| f_\text{OTE}^{+}|$ (a) and $| f_\text{ETO}^{+}|$ (b), given in \eref{eq:FLR}. 
Both density plots show that pairing is enhanced on resonance, i.e., at energies roughly around the bound states localized inside the double cavity (cf. \fref{fig:nonlocal}). 
Equal-spin $\up\up$ pairing is forbidden within the left lead, while it connects electrons and holes from different leads via CAR processes. 

OTE and ETO components are complementary and add up to a total spin $\up\up$ contribution, which is closely tied to the bound state energies. We show in \fref{fig:results1}(c) a cut of the density plots along the interface between the rightmost F region and normal electrode ($x \! = \! 6.5\xi_0$). The density of states is shown in black dashed lines and clearly envelops the contributions from the pairing. 
This explicitly reveals the connection between the local density of states at each cavity and the nonlocal induced pairing between cavities, owing to the bound states present in the system. 

MBS and ABS formed in the cavities between S and F barriers behave in a very different manner. The nonlocal pairing amplitudes plotted in \fref{fig:results1}(d,e) show a different content of even and odd-frequency pairing for MBS compared to ABS. 
First, for energies associated with hybridized MBS, \textit{cf}. \fref{fig:results1}(d), the pairing is almost constant within the cavities and in the right lead. The pairing at higher energy due to ABS resonances, contrarily, features stronger variations in both energy and space, see \fref{fig:results1}(e). 
This effect is clearly explained by \eref{eq:FLR}, since for $\omega\!\rightarrow\!0$ the spatial dependence becomes an irrelevant global phase factor. 
Additionally, the ETO and OTE content of MBS and ABS is also strikingly different. For MBS, odd-frequency pairing increases as we take the spatial coordinate further away from the leftmost lead, reaching a maximum for the nonlocal paring between cavities. Interestingly, OTE is constant and clearly dominant over ETO inside the rightmost region, making it the most important contribution to CAR processes at low energies\cite{Crepin_2015}. ABS, on the other hand, feature an oscillatory spatial behavior where ETO and OTE pairing amplitudes alternate.  

For the validity of the previous results, it is crucial that the states between cavities are hybridized. In the absence of hybridization, i.e., when the hybridization phase in \fref{fig:hyb} reaches a multiple of $\pi$ causing destructive interference, the nonlocal pairing vanishes, \textit{cf}. \fref{fig:hyb}. 
It is important to notice that the nonlocal pairings in \eref{eq:FLR} vanish because the scattering amplitudes $t_\mathit{hL,eR}$ and $t_\mathit{hR,eL}$ are zero. Therefore, CAR processes are also forbidden in the absence of hybridization. 
Once more, the nonlocal pairing is reflected in the local density of states, plotted in \fref{fig:hyb} as a black dashed line. Indeed, hybridized MBS are accompanied by dominant OTE pairing as long as the hybridization phase is not too close to $\pi$. On the other hand, hybridized ABS feature comparable OTE and ETO contributions for all hybridizations.

The previous results are valid for a wide range of parameters beyond the choice in \fref{fig:results1} and \fref{fig:hyb}. One of our main results is that the hybridized and non-hybridized cavities behave in a very different manner. For any choice of parameters, one just needs to determine if the corresponding hybridization phase results in destructive interference or not. 
For clarity, we have only chosen symmetric junctions with equal cavity widths. 
An asymmetric setup would only affect the pairing at ABS resonances. Since in this case the energies of ABS in the left and right cavity do not match, one cannot have degenerate states and perfect destructive interference. CAR amplitude and nonlocal pairing will thus remain a finite superposition of ETO and OTE contributions around ABS resonances. The behavior of the pairing for the MBS resonances, however, remains unimpaired. This is captured in \eref{eq:sf-cav}, where the effect of the asymmetry is reduced as $\omega\!\rightarrow\!0$. 

Furthermore, the close connection between equal-spin pairing and the bound states between S and F barriers implies that our findings generalize to setups with a central F region flanked by two superconductors. We have shown in \sref{sec:hyb} that the hybridization parameter --which would now have a different physical origin-- has the exact same effect on the behavior of the bound states. This ultimately translates in analogous behavior of the nonlocal, equal-spin pairing amplitudes as well. Local pairing amplitudes within the same lead, however, would in fact be different, since they depend on local normal or Andreev reflection. 
Nevertheless, these are necessarily singlet or zero $z$-component triplet amplitudes. 

\begin{figure}
\includegraphics[width=\columnwidth]{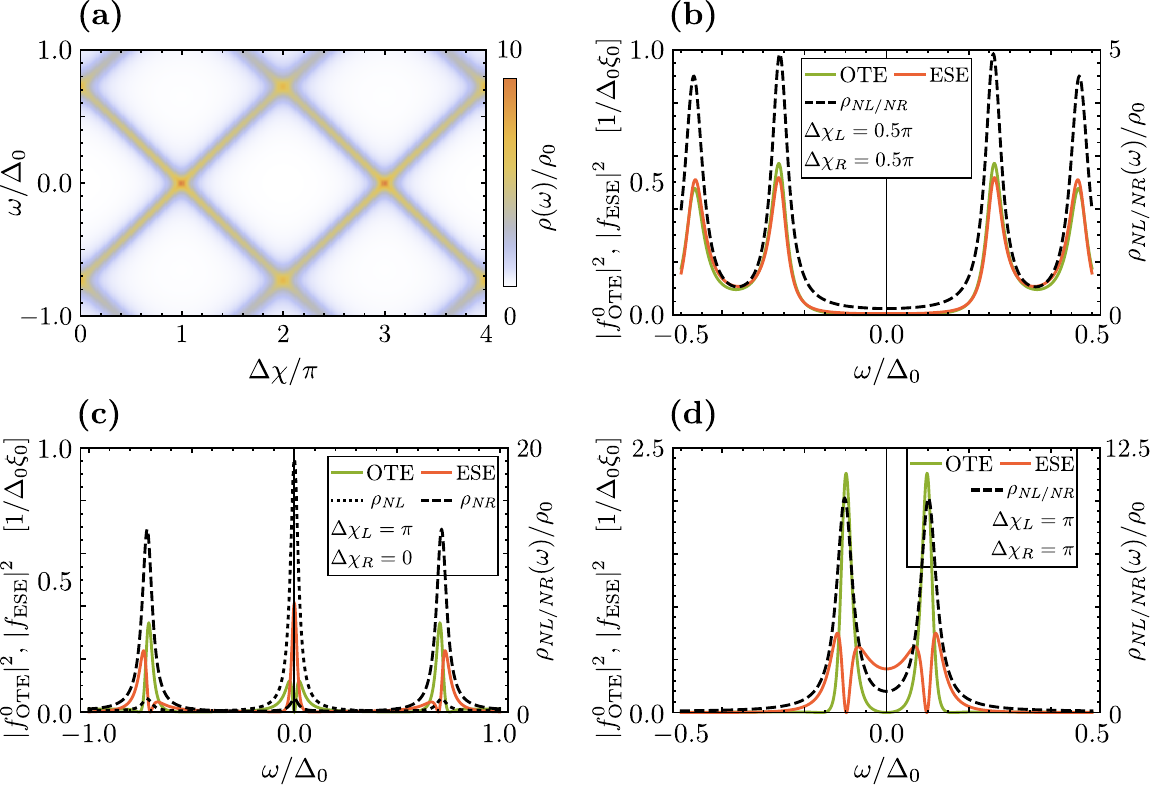}
	\caption{\label{fig:results2a} 
		LDOS and pairing amplitude for setups with only superconductors. 
        (a) Energy levels of the S-S cavity as a function of the phase difference between superconductors. 
        (b) - (d) For the S-S-S double cavity, nonlocal odd-frequency triplet and even-frequency singlet pairing amplitudes between the left lead L and the intermediate normal region NR ($x'\!=\!-\xi_0$ and $x\!=\!4.5\xi_0$). The dotted/dashed black lines show the LDOS. The phase differences are (b) $\Delta\chi_L\!=\!\Delta\chi_R\!=\!0.5\pi$, (c) $\Delta\chi_L\!=\!\pi$, $\Delta\chi_R\!=\!0$, and (d) $\Delta\chi_L\!=\!\Delta\chi_R\!=\!\pi$. For all plots we use the widths $d_{\mathit{NL}}\!=\!d_{\mathit{NR}}\!=\!1.25\xi_0$, $d_{\mathit{SL}}\!=\!d_{\mathit{SM}}\!=\!d_{\mathit{SR}}\!=\!1.5\xi_0$, and $\mu\!=\!\Delta_0$ in every region. 
		}
\end{figure}
\begin{figure}
\includegraphics[width=\columnwidth]{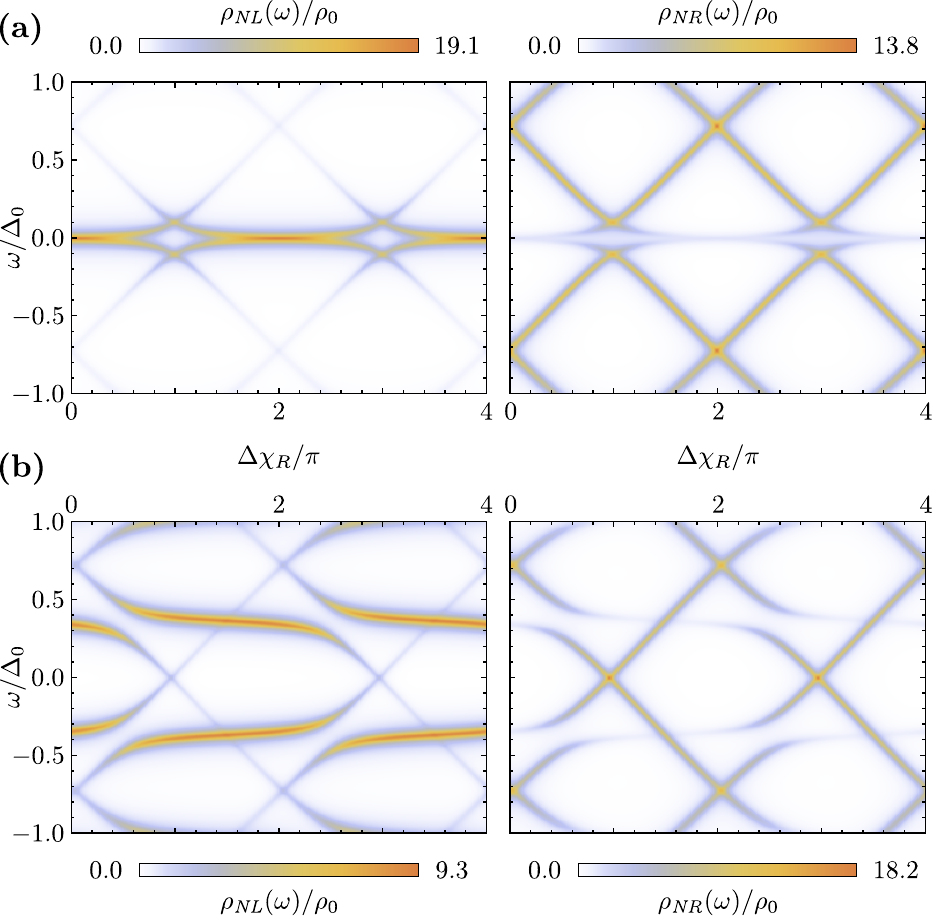}
\caption{\label{fig:results2} 
	Local density of states for the setup with three superconductors (S-S-S) for $\Delta\chi_L\!=\!\pi$ (top row) and $\Delta\chi_L\!=\!0.5\pi$ (bottom row). The LDOS is calculated in the left cavity (right cavity) in left (right) column. 
        }
\end{figure}

\subsection{SSS setup \label{sec:sss}}  
We now consider the case of a double cavity formed by superconductors only. 
This setup is intrinsically different from the previous cases, but is easier to achieve experimentally since it does not require ferromagnetic elements. 
In the S-S cavity, electrons and holes with opposite spins form Andreev bound states. The lack of F regions implies that the ABS formed from right-moving spin up electrons do not couple to the ones formed from left moving spin-down electrons. As a result, the ABS in the cavity are spin-degenerate and have topologically protected zero-energy crossings\cite{FuKane_2009} when the phase difference between superconductors is $\Delta\chi\!=\!n\pi$, with $n$ an integer [see \fref{fig:results2a}(a)]. 
The analysis of the pairing amplitude in the S-S cavity reveals that the ABS are a superposition of singlet and triplet states, where the triplet component is not spin polarized although it features both ETO and OTE terms\cite{Cayao_2017}. 

The double S-S-S cavity strongly depends on the relative phase differences between the middle (C) and outer superconductors (L and R), i.e., $\Delta\chi_{L}\!=\!\chi_{M}\!-\!\chi_L$ and $\Delta\chi_{R}\!=\!\chi_{R}\!-\!\chi_{M}$. 
By placing two cavities together, the dependence of the ABS on the phases becomes more intricate. 
In \fref{fig:results2a}(b,c,d) we plot the LDOS in both cavities as a function of the energy for different values of $\Delta\chi_{L,R}$ (black dotted and dashed lines). 
Similarly to the single cavity ABS, and contrary to the previous F-S-F cavities, the ABS are not confined to a specific set of energies, but evolve inside the gap as a function of the relative phase differences. 
The ABS patterns from each cavity are completely decoupled in \fref{fig:results2a}(b), featuring four resonances. Two of the levels can combine at zero energy as shown in \fref{fig:results2a}(c). When the total phase difference across each cavity is $\pi$, the two patterns coincide again, cf. \fref{fig:results2a}(d). 

We also include in \fref{fig:results2a}(b,c,d) the ESE and OTE components of the pairing. We notice that even frequency terms ESE and ETO have the same energy dependence and only differ in magnitude; the same applies to OSO and OTE. Thus, we only show ESE and OTE terms for simplicity. 
All panels illustrate the nonlocal pairing between the leftmost region and the right cavity. Even and odd frequency terms have similar magnitudes and are mingled along the energies of the ABS. 
Even-frequency terms dominate for states at zero energy, where the odd-frequency terms naturally vanish. The opposite situation arises for finite but small energies, cf. \fref{fig:results2a}(c) and \fref{fig:results2a}(d). As the energy of the ABS moves closer to the gap edges, the magnitude of even and odd-frequency terms becomes comparable. 

Interestingly, the hybridization of the ABS in this setup cannot be completely removed. We test this statement in a simple S-S-S setup where the outer regions are semi-infinite superconductors and both cavities have the same width $d$. We analyze the common denominator of the scattering amplitudes for the injection of electrons from the leftmost region. This quantity is proportional to the denominator of the Green function, so its zeros provide the spectrum of the junction. We find that the denominator contains two terms, namely, 
\begin{equation}\label{eq:sss-cav}
D_1+\e^{2\kappa L}D_2 = D_1+D_2+\left(\e^{2\kappa L}-1\right)D_2 ,
\end{equation}
with $\kappa\!=\!\sqrt{\Delta_0^2-\omega^2}$ and
\begin{align*}
D_1={}& - \sin\left(d \omega + \Delta\chi_{L}/2 \right) \sin\left(d \omega + \Delta\chi_{R}/2 \right)  , \\ 
D_2={}&  \sin\left( d\omega-\eta+\Delta\chi_L/2 \right) \sin\left( d\omega-\eta+ \Delta\chi_R/2 \right) ,
\end{align*}
where $\eta\!=\!\arccos(\omega/\Delta_0)$. 
$D_2$ is the product of two resonance conditions for independent cavities, \textit{cf}. \eref{eq:ss-cav}. 
For $L\!=\!0$, the denominator reduces to
\begin{equation*}
D_1 + D_2 = \sin(\eta) \, \sin\left[ 2d\omega - \eta + (\Delta\chi_L + \Delta\chi_R)/2 \right] ,
\end{equation*}
which is equivalent to the condition for the bound states of an S-S cavity of width $2d$ and relative phase $\Delta\chi_L\!+\!\Delta\chi_R \!=\! \chi_R \!-\! \chi_L$. 
The rightmost term in \eref{eq:sss-cav} interpolates between this case and the situation with two independent cavities of width $d$, obtained in the limit $L \!\rightarrow\! \infty$ (the divergence in \eref{eq:sss-cav} is countered by the numerator). 
We immediately see that there is no zero-energy state for $\Delta\chi_L\!=\!\pi$ and $\Delta\chi_R\!=\!\pi$, since $D_1\!+\!D_2\!\neq\!0$ for finite $L$. 

A zero energy solution ($\omega\!=\!0$) can still be found localized in one of the cavities for the conditions $\Delta\chi_{L}\!=\!\pi$, $\Delta\chi_{R}\!=\!0$ (left cavity) or $\Delta\chi_{L}\!=\!0$, $\Delta\chi_{R}\!=\!\pi$ (right cavity), see the two panels of \fref{fig:results2}(a). Although both conditions fulfill that $D_1\!+\!D_2\!=\!0$, the numerator of the Green function suppresses one the resonance in one of the cavities, cf. right panel of \fref{fig:results2}(a). 
A similar analysis can be done for the scattering process consisting of injected spin-down electrons from the rightmost region and the results would be the same, up to an overall sign in the phase differences. Therefore, the ABS remain uncoupled and the LDOS exhibits the superposition of two different patterns as shown in \fref{fig:results2}(b). 


\section{Conclusion \label{sec:conc}}
We have analyzed the hybridization of Majorana bound states at the helical edge of quantum spin Hall insulators. In our theoretical model, confinement is achieved by two different means: (i) proximity to ferromagnetic insulators (F) and (ii) proximity to $s$-wave superconductors (S). We have looked at certain combinations of those barriers, i.e., F-S-F, S-F-S, and S-S-S setups. Each of these combinations can be viewed as a ``double-dot'' structure for Majorana and Andreev bound states at the quantum spin Hall edge. Interestingly, we have identified different ways to tune the hybridization between the bound states in each of the ``dots'' by playing with system parameters, e.g., the relative angle between the magnetization directions of two F regions or the superconducting phase difference between two S regions. This understanding could be useful for applications in topological quantum computing.

Furthermore, we have shown how the bound states are connected to unconventional superconducting pairing. This pairing could be of triplet type and becomes spin polarized for junctions including ferromagnets. It also shows odd-frequency behavior under generic conditions. 
We have put emphasis on the nonlocal (in real space) pairing because of its rich information content in our system due to the helicity of particle and hole excitations. Remarkably, we have been able to identify a strong connection between the nonlocal pairing amplitude and the local density of states. We have come full circle because the bound states in the ``double-dot'' structure are indeed responsible for this unusual connection between nonlocal characteristics of a system (the pairing amplitude) and a local observable (the density of states).

In the future, we would like to extend this analysis to interacting systems in which Majorana fermions can become parafermions. To the best of our knowledge, the relation of parafermions to unconventional superconducting pairing is not understood at all.  


\acknowledgments 
We thank M. Stehno and Y. Tanaka for fruitful discussions. 
Financial support by the DFG (SPP1666 and SFB1170 ``ToCoTronics'') and the ENB Graduate School on ``Topological Insulators'' is acknowledged. 
P.B. acknowledges funding from the European Union's Horizon 2020 research and innovation programme under the Marie Sk\l odowska-Curie grant agreement No. 743884. 



\appendix

\section{Symmetry analysis \label{app:sym-f}}
Superconducting correlations in the system are described by the causal anomalous Green function
\begin{align}\label{eq:causal}
\mbf{F}^{++}_{\alpha\beta,\sigma\sigma'}\left(\tau\right) ={}& -i\theta\left(\tau\right) \mean{ \mbf{c}_{\alpha\sigma}\left(\tau\right) \mbf{c}_{\beta\sigma'}\left(0\right) } \nonumber \\ & + i\theta\left(-\tau\right) \mean{ \mbf{c}_{\beta\sigma'}\left(-\tau\right) \mbf{c}_{\alpha\sigma}\left(0\right) } , 
\end{align}
with $\tau\!=\!t\!-\!t'$ the difference between the time coordinates of the fermions forming the Cooper pair, which are described by the annihilation operators $\mbf{c}_{\alpha\sigma}\left(\tau\right)$. 
In the following, the indices $\sigma,\sigma'$ correspond to the spin degree of freedom, and $\alpha,\beta$ label the position coordinates. We assume that the rest of quantum numbers do not play any role. 
Analogously, the anti-causal anomalous Green function, defined on the negative branch of the Keldysh contour, is given by
\begin{align}\label{eq:anticausal}
\mbf{F}^{--}_{\alpha\beta,\sigma\sigma'}\left(\tau\right) ={}& i\theta\left(\tau\right) \mean{ \mbf{c}_{\beta\sigma'}\left(0\right) \mbf{c}_{\alpha\sigma}\left(\tau\right) } \nonumber \\ & - i\theta\left(-\tau\right) \mean{ \mbf{c}_{\alpha\sigma}\left(0\right) \mbf{c}_{\beta\sigma'}\left(-\tau\right) } . 
\end{align}

To illustrate the possible creation of Cooper pairs in our system, we set two positions $L$ and $R>L$ and we consider a symmetric superposition of the causal Green functions with both spatial coordinates. Taking the limit $\tau\!\rightarrow\!0$, we find
\begin{align}
\begin{gathered}\label{eq:causal-LR}
\mbf{F}^{++}_{(LR+RL),\sigma\sigma'}\left(0\right) = \\
-i\mean{ \mbf{c}_{R\sigma'}\left(0^+\right) \mbf{c}_{L\sigma}\left(0\right) } -
\underbrace{i\mean{ \mbf{c}_{L\sigma}\left(0^+\right) \mbf{c}_{R\sigma'}\left(0\right) } }_\text{forbidden by helicity if $\sigma=\sigma'$}
, 
\end{gathered}
\\ 
\begin{gathered}\label{eq:anticausal-LR} 
\mbf{F}^{--}_{(LR+RL),\sigma\sigma'}\left(0\right) = \\
i\mean{\mbf{c}_{L\sigma}\left(0\right)  \mbf{c}_{R\sigma'}\left(0^+\right) } +
\underbrace{i\mean{ \mbf{c}_{R\sigma'}\left(0\right) \mbf{c}_{L\sigma}\left(0^+\right)  } }_\text{forbidden by helicity if $\sigma=\sigma'$} . 
\end{gathered}
\end{align}
\eref{eq:causal-LR} indicates two possible ways to create a Cooper pair: an electron is annihilated at position $L$ or $R$ after another electron was destroyed at a previous time in position $R$ or $L$, respectively. Notice that the anti-causal anomalous Green function describes the same processes \textit{for negative relative times}. 

The general result of \eref{eq:causal-LR} changes drastically when we consider that our system is the helical edge of a QSHI. In this case, if $\sigma=\sigma'$, the latter terms of \eref{eq:causal-LR} and \eref{eq:anticausal-LR} are forbidden by helicity. By restricting the propagation of equal spin particles to only one direction, we have effectively excluded one of the possible ways for creating Cooper pairs. 

We can recover this ``lost symmetry'' by realizing that the surviving term of the anti-causal anomalous Green function corresponds to the forbidden term of the causal function for negative relative times [\textit{cf}. rightmost term of \eref{eq:causal}]. Consequently, a combination of causal and anti-causal anomalous Green functions should recover the missing symmetry of the system, that is,
\begin{equation}  \label{eq:GF-sym}
\begin{gathered}
\mbf{F}^{++}_{(LR+RL),\sigma\sigma}\left(0\right) - \mbf{F}^{--}_{(LR+RL),\sigma\sigma}\left(0\right)  \\ = 
-i\mean{ \mbf{c}_{R\sigma}\left(0^+\right) \mbf{c}_{L\sigma}\left(0\right) } 
-i\mean{ \mbf{c}_{L\sigma}\left(0^+\right) \mbf{c}_{R\sigma}\left(0\right) } . 
\end{gathered}
\end{equation}
We immediately see that this result is equivalent to superposing the parts with $\tau>0$ and $\tau<0$ in the definition of the spatially symmetric causal Green function, excluding the terms forbidden by helicity [\textit{cf}. \eref{eq:causal}]. 
Moreover, the full causal and anti-causal Green functions in Nambu space are connected to the retarded and advanced ones as $\mbf{G}^{++}\!-\!\mbf{G}^{--}\!=\! \mbf{G}^R \!+\! \mbf{G}^A$. 
As a result, the anomalous part of this Green function is odd under the exchange of the two fermion operators,
\begin{equation}\label{eq:GF-sym2}
\mbf{F}^R_{\alpha\beta}\left(\tau\right) + \mbf{F}^A_{\alpha\beta}\left(\tau\right) = - \left[\mbf{F}^R_{\beta\alpha}\left(-\tau\right) + \mbf{F}^A_{\beta\alpha}\left(-\tau\right) \right] ,
\end{equation}
thus fulfilling the general requirement of Fermi-Dirac statistics. 
In brief, due to the restrictions imposed by helical systems, neither the causal nor the anti-causal Green's functions fulfill the typical symmetries that stem from Fermi-Dirac statistics themselves (the same applies to retarded and advanced functions). The combination of both does it, however. It is thus the object of interest for the symmetry classification of pairing amplitudes of helical systems. 

We now focus on the symmetry classification of the anomalous Green function $\mbf{F}^R+\mbf{F}^A$, defined in \eref{eq:GF-sym}. We make use of the connection between the full retarded and advanced Green functions in Nambu space and their transformation under particle-hole symmetry (PHS), namely
\begin{align}\label{eq:syms}
\mbf{G}^A_{\alpha\beta,\sigma\sigma'}\left(\omega\right) ={}& \left[\mbf{G}^{R}_{\beta\alpha,\sigma'\sigma}\left(\omega\right) \right]^* , \\  \mbf{P}\mbf{G}^{R,A}\left(\omega\right)\mbf{P}^{-1}={}& -\mbf{G}^{R,A}\left(-\omega\right) ,
\end{align}
where $\mbf{P}\!=\!\sy\ty\mbf{K}$, with $\mbf{K}$ the complex conjugation operator, is the PHS operator. In the basis $\mbf{\Psi}=(\mbf{c}_{\up},\mbf{c}_{\dw},\mbf{c}_{\up}^{\dagger},\mbf{c}_{\dw}^{\dagger})^T$, the PHS operator is given by $\mbf{P}=\si\tx\mbf{K}$. 
Combining these two properties, we find
\begin{equation*}
\mbf{F}^A_{\alpha\beta,\sigma\sigma'}\left(\omega\right) = -\mbf{F}^{R}_{\beta\alpha,\sigma'\sigma}\left(-\omega\right) ,
\end{equation*}
where $\mbf{F}$ is the anomalous Green function corresponding to the electron-hole Nambu component of $\mbf{G}$. 

Imposing the symmetric and anti-symmetric superposition of the spin indices, we find the spin-singlet scalar and the spin-triplet pseudo-vector anomalous Green functions
\begin{equation*}
\begin{gathered}
\mbf{F}^{R}_{\alpha\beta,\mu}\left(\omega\right) = \\
\left\{ \begin{array}{l}
\mbf{F}^{R}_{\alpha\beta,0}\left(\omega\right) \equiv \mbf{F}^{R}_{\alpha\beta,(\up\dw-\dw\up)}\left(\omega\right) = \mbf{F}^{R}_{\alpha\beta,\up\dw} - \mbf{F}^{R}_{\alpha\beta,\dw\up} \\ 
\mbf{F}^{R}_{\alpha\beta,3}\left(\omega\right) \equiv \mbf{F}^{R}_{\alpha\beta,(\up\dw+\dw\up)}\left(\omega\right) = \mbf{F}^{R}_{\alpha\beta,\up\dw} + \mbf{F}^{R}_{\alpha\beta,\dw\up} \\
\mbf{F}^{R}_{\alpha\beta,+}\left(\omega\right) \equiv \mbf{F}^{R}_{\alpha\beta,\up\up}\left(\omega\right) \\
\mbf{F}^{R}_{\alpha\beta,-}\left(\omega\right) \equiv \mbf{F}^{R}_{\alpha\beta,\dw\dw}\left(\omega\right) 
\end{array} \right. , 
\end{gathered}
\end{equation*}
with $\mu=0,+,-,3$. We define the exchange of spin index $\bar{\mu}$ so that $\mbf{F}^{R}_{\alpha\beta,\bar{\mu}}\!=\!(-\mbf{F}^{R}_{\alpha\beta,0},\mbf{F}^{R}_{\alpha\beta,j})$, with $j=+,-,3$. 

Analogously, we can construct the anomalous Green function that is symmetric or anti-symmetric with respect to the spatial coordinates, defining
\begin{equation*}
\mbf{F}^{R,A}_{\pm,\mu}\left(\omega\right) \equiv \mbf{F}^{R,A}_{(\alpha\beta\pm\beta\alpha),\mu}\left(\omega\right)= \mbf{F}^{R,A}_{\alpha\beta,\mu}\left(\omega\right) \pm \mbf{F}^{R,A}_{\beta\alpha,\mu}\left(\omega\right) .
\end{equation*} 
From the advanced anomalous Green function, symmetric both in spin and real space indices, we find
\begin{align*}
\mbf{F}^{A}_{\pm,\mu}\left(\omega\right) ={}& \mbf{F}^{A}_{\alpha\beta,\mu}\left(\omega\right) \pm \mbf{F}^{A}_{\beta\alpha,\mu}\left(\omega\right) \\ ={}&
-\mbf{F}^{R}_{\beta\alpha,\bar{\mu}}\left(-\omega\right) \mp \mbf{F}^{R}_{\alpha\beta,\bar{\mu}}\left(-\omega\right) = \mp \mbf{F}^{R}_{\pm,\bar{\mu}}\left(-\omega\right) . \nonumber
\end{align*} 

Going back to the time-symmetric anomalous Green function $\mbf{F}^R+\mbf{F}^A$, we can write
\begin{align*}
\mbf{F}^{R}_{\pm,\mu}\left(\omega\right) + \mbf{F}^{A}_{\pm,\mu}\left(\omega\right) ={}& 
\mbf{F}^{R}_{\pm,\mu}\left(\omega\right) \mp \mbf{F}^{R}_{\pm,\bar{\mu}}\left(-\omega\right) \\ ={}& 
\left\{ \begin{array}{c}
\mbf{F}^{R}_{\pm,0}\left(\omega\right) \pm \mbf{F}^{R}_{\pm,0}\left(-\omega\right) \\ 
\mbf{F}^{R}_{\pm,i}\left(\omega\right) \mp \mbf{F}^{R}_{\pm,i}\left(-\omega\right) 
\end{array} \right. . \nonumber
\end{align*}
We immediately see that the singlet component that is  (anti-) symmetric in spatial coordinates is (odd) even in frequency. They correspond to ESE and OSO symmetry classes. For the triplet components, the spatial symmetric and anti-symmetric combinations are odd and even in frequency, respectively. They belong to the OTE and ETO classes. 


\bibliography{2dTInonlocal}

\end{document}